# Dynamic Link adaptation Based on Coexistence-Fingerprint Detection for WSN

Charbel Nicolas and Michel Marot
CNRS-SAMOVAR-UMR 5157, Département-RST
Institut TELECOM:TELECOM SudParis
9 rue Charles Fourier, 91011 Evry CEDEX, France
{Charbel.Nicolas, Michel.Marot}@telecom-sudparis.eu

*Abstract*—Operating in the ISM band, the wireless sensor network (WSN) risks being interfered by other concurrent networks. Our concerns are the technologies that do not perform listening before transmission such as Bluetooth, and the ones that do not detect other technologies due to their channel sensing techniques like WIFI. To overcome this issue a WSN node should be able to identify the presence of such technologies. This will allow deducing the characteristics of the generated traffic of these technologies, and thus the behavior of the channel can be predicted. These predictions would help to trigger adequate reactions as to avoid or synchronize with the concurrent networks. Many works exist on link adaptation, but they concern blind adaptations which are unintelligent and solve momentarily the problem that may reappear over time.

In this paper, we perform several experiments on a real testbed to categorize the model of the bit errors in corrupted received packets. These experiments are performed under different conditions of channel noise and interferences. This allows us to identify each corruption pattern as a fingerprint for the interfering technology. Then we propose the mechanism FIM to identify on the fly the source of the corruption. With an implementation on "Tmote Sky" motes using Tinyos1.x, We demonstrate the use of FIM for link adaptation in a coexistence environment. Our mechanism led to throughput improvements of 87%-100% depending on the transmission rate and channel quality.

## I. INTRODUCTION

IEEE 802.15.4 used by Zigbee and WSNs represents one of the most used and deployed network types in the industrial and medical environments. In this paper, Zigbee and WSN implicitly refer to IEEE802.15.4. These WSNs share the 2.4 GHz industrial, scientific, and medical (ISM) band with other technologies like Bluetooth or IEEE 802.11b/g (WLANs). Due to the coexistence at the same ISM band, without any radio planning, the interference on the WSN by these technologies is inevitable. This is a key issue since WSNs are very sensitive to the environment (cf. [1]) and they often use network performance measurements to feed configuration algorithms. The goal of this paper is to design an adaptive mechanism allowing the WSN nodes to dynamically recognize the presence of alternative technologies and to adapt their transmission modes accordingly.

The source of this coexistence vulnerability is the heterogeneity of the physical and of the MAC mechanisms that leads to collisions or interferences. The coexistence with WiFi has a big effect on the WSN performance, especially on collision avoidance and fairness between both protocols. This effect is primarily characterized by the WiFi and Zigbee transmission (Tx) rates. In contrast, our work and also [2] show that Bluetooth interference, thanks to its frequency hopping nature, has a smaller effect on the WSN. The smaller percentage of packet loss comes from the fact that Bluetooth does not use channel sensing mechanism but instead employs FH/TDD for channel access.

Due to its low power and low bit rate, the WSN is mostly affected by the technologies that have transmission powers and medium sensing thresholds higher than its own. The most common MAC layer protocol used in these wireless communications is the CSMA/CA. When it was first developed, the diversity of the technologies such as IEEE802.11 and IEEE802.15.4 did not exist. By adapting the CSMA/CA to these technologies, modifications have been introduced on its sensitivity thresholds, Tx rates, etc. This led, in a coexistence environment, to the loss of the fairness and collision avoidance initially provided in CSMA/CA.

The deficiency cause in CSMA/CA is the clear channel assessment (CCA) techniques. The different types of existing channel sensing methods are (cf. [3]): the Energy Detection (ED), Preamble Detection (PD) and Decorrelation-Based CCA (DB). Alternative spectrum sensing methods used in cognitive radio including multitaper spectral estimation, wavelet transform based estimation, Hough transform, and time-frequency analysis cannot be applied in IEEE802.15.4 due to limitations in executing complex calculations and to energy constraints. PD cannot be used by IEEE802.15.4 to detect IEEE802.11 transmission because of the energy cost and processing complications (the need of high sampling rate, filtering, etc). DB is a combination of ED and PD, which inherits by that the same limitations. ED is not very efficient in case there are wideband signals, the transmission power being near the noise threshold with the spectrum spreading technique (cf. [3]) and therefore it is difficult to detect the transmission. Moreover, depending on the used channels, the ZigBee channel may overlap the WiFi one. The ZigBee channel is 2MHz wide (cf. Section 3) and the WiFi one is 22MHz [2]. Because of the smaller emission power used by ZigBee and an averaging of the energy received over the WiFi 22Mhz bandwidth, a WiFi device cannot always detect a ZigBee transmission.

The inadequacy of CSMA/CA in heterogeneous environment and the inability to determine the cause of corruptions lead the link layer protocols designed for WSN to react blindly to a corrupted packet. Not knowing the nature of the concurrent network leads to the excessive channel power sampling to exploit white spaces that occur when there is coexistence with other technologies. To better exploit and predict white spaces in a channel, it is necessary to determine the technology that occupies the channel, and based on its characteristics white spaces can be exploited.

If the coexistence cannot be avoided, corrupted packets are necessarily received and the network must adapt intelligently. The link adaptation can exist on the routing level by changing the network model or on the MAC and physical layer by changing the transmission power, channel, rate, adding redundant bits, etc. In a self organized network where the energy is a limited resource, analyzing the quality and stability of the channel is an energy and throughput costly process. We claim that the exact cause of the packet error can be inferred by simply analyzing the errors in the received packets and then the best countermeasure can be chosen. Since packet corruptions are unavoidable, and if a repetition mechanism like A.R.Q. is used, the correct packet is finally received. Then, instead of ignoring the previously received corrupted packet occurrences, we suggest to get them, to compare them with the finally received correct packet in order to detect the shape of the error sequences and to use this information to infer the cause of the errors. Each cause should produce a different fingerprint.

Knowing the cause of the erroneous packets can help to take adequate decisions at the link layer. By identifying a WiFi fingerprint in the corrupted bytes of the Zigbee packets, a channel swap can be made. If Bluetooth is detected, smaller packet lengths can be used. In case of weak link detection, redundant bits or FEC can be applied. If hidden terminal are recognized, RTS/CTS or resynchronization (if the network uses TDMA) must be applied. Decisions can also be taken at the routing layer. For clustered network, the cluster head selection may be based on its surrounding technologies and their effect on its communication. When a tree based topology is built, the nodes under permanent interferences with a different technology may be chosen as leaf nodes. In ad-hoc TDMA based network, the time slot that is interfered by other technologies may be black listed.

In this paper, we develop this idea from experimentations made using off-the-shelf sensor devices. Our main contributions are: 1) an empirical study on byte error patterns for different error causes, 2) the identification of the error patterns as fingerprints to specific technologies, 3) the proposal of a new mechanism, Fingerprint Identification Mechanism (FIM) as a light weight and energy efficient mechanism to identify on the fly the coexisting technologies based on their fingerprints, its application for a link adaptation mechanism and, finally,4) the implementation and evaluation of FIM on a real WSN testbed. Our work is useful for a lot of applications. Specifically in the applications that need a high packet transmission rate used in intrusion detection, cold chain monitoring and health monitoring.

In the next part the related work is discussed, then the analysis carried to identify the different byte error patterns in different contexts are presented. The last part is dedicated to the fingerprint identification mechanism, its reactions to error situations and finally its performance evaluation.

## II. RELATED WORK

Different protocols have been developed to adapt the IEEE802.15.4 nodes in case of coexistence with networks using other technologies. Most of these protocols are focused on WiFi because of its high impact on the Zigbee performance. Their deficiency is that each protocol is not developed to detect the coexistence but they assume that a Zigbee network is coexisting with WiFi, and then they try to adapt ZigBee accordingly. For example, the CCA threshold of Zigbee is adjusted to minimize the packet loss in [4] and [5]. Others modify the packet length based on a WiFi white space model as in [6].

Work determining a coexistence case has been based on power (RSSI) sampling. In [7], energy detection spectrum sensing to select best channel in the presence of WiFi and Bluetooth is proposed. The power spectrum density of WiFi and Bluetooth is analyzed by using a sensor. In [8], RSSI sampling is used to detect the periodic beacons of WiFi. However, these solutions based on power sampling suffer from at least one of the following issues. 1) They are based on detecting the power of the physical channel and thus the Zigbee transmission power occupying the channel cannot be differentiated from WiFi or Bluetooth transmission power. 2) A high sampling rate is needed to detect WiFi packets, at least two times higher than the duration of a packet, which requires a high energy cost and long processing times in Zigbee nodes. 3) These approaches need a dedicated time to do the sampling and to converge.

To our knowledge, none of the work dealing with coexistence adaptation or detection is based on the captured corrupted packets.

## III. EMPIRICAL MEASUREMENTS AND ERROR PATTERN IDENTIFICATION

### 1) Used equipments and topology

The aim of the experiments is to characterize the statistical corruption patterns in a Zigbee packet, relatively to each coexisting technology. Our testbed consists of 802.15.4-compliant Tmote-Sky motes equipped with Chipcon CC2420 transceiver [11], in presence of an interfering signal originating from:

a. Intel ® pro/wireless LAN 2100 3A Mini PCI adapter and intel ® WIFI link 5100 AGN (WIFI cards integrated to DELL laptops).
b. The built-in Bluetooth of Dell latitude (the receiver) and the Bluetooth of NOKIA E51 (the transmitter) allowing to send a long video file at the rate of 2.1Mbps. The Bluetooth transmitter uses TDM and FHSS, 79 channels each one being 1MHz wide. There are 1600 hops per second (hps), the time slot is thus 625μs. Then the equivalent rate transmitted by Bluetooth on a 2MHz wide ZigBee channel is 2MHz/(79x1MHz)x2.1Mbps=53kbps.

To monitor the channel characteristics during the tests, to tune and select the overlapping channels between WiFi and Zigbee, we used the BK PRECISION by MICRONIX (8.5GHz) 2658 spectrum analyzer. We specifically used the channel 1 for WiFi and the channels 11, 13 and 14 for Zigbee. The Tx power of Zigbee was set to -7dBm (PA_LEVEL=15).

Operating in the 2.4GHz band IEEE802.15.4 forced the use WIFI versions IEEE802.11b/g. Moreover, between the IEEE802.11b and IEEE802.11g, there is an important aspect to be taken which is that the IEEE802.11g uses the OFDM modulation that applies low CCA threshold. This fact allows the IEEE802.11g to detect the presence of IEEE802.15.4 transmissions and apply back-off. In contrast, IEEE802.11b uses DSSS (high Tx power) therefore it uses higher thresholds. As we are dealing with packet corruption due to hidden terminal effect and not congestion, this study is focused on IEEE802.11b.

To speed up the convergence, we submitted the equipment to extreme conditions and very high Tx rates. For these experiments, we used Tinyos 1.x to modify the TOSBASE application to receive and analyze packets. In order to receive erroneous packets, we deactivated the CRC check and the hardware acknowledgments generated by the CC2420 [11]. To have a high collision rate and consequently a fast convergence rate, we disabled the CCA and backoff (they are reactivated later for the FIM validation) and we set the appropriate channels and power to be used. We used the APIs developed in [9] to generate a customized WiFi traffic. To monitor the WiFi traffic (control and Data packets), we used Wireshark. The collision analysis is done on a fixed Zigbee packet length of 122bytes for all experiments.

*2) Measurement Results*

In this part, collisions are generated to determine the technology fingerprints. The measured quantity is the distribution of the number of erroneous bytes in a received packet. We run the experiments within 5 scenarios corresponding to the most typical contexts of errors. In all the experiments, the influence of the packet rates, packet sizes and channel numbers is observed. When collisions occur, the Zigbee packet we analyze is (a) not received or (b) it is received with an error or (c) it is received correctly. The case (a) occurs when the collision corrupts the synchronization header (SHR) and the Frame length. During all the tests, we directly dropped the packets with corrupted length because it causes the applications to malfunction. The scenarios are:

- **The case of Weak SNR for ZigBee Tx**

Long distances and low Tx powers create weak links. We ran an experiment with a low emission power. For the weak SNR test represented in Figure 1, 105849 packets were transmitted of which 18% suffered from corruption. The transmitters and the receivers were placed in different rooms separated by a distance of 5 meters. To achieve solid results, the experiments involved all the channels. Due to the lack of space and future need we only show the results of the channels 11, 13 & 14. For all the channels, we obtained similar results. These channels being theoretically quasi-orthogonal, there is practically no interference between adjacent nodes using co-channels and the same power for packet transmission: each traffic experiences no collision. Special measures have been taken to ensure that no interfering transmitters were present, thus avoiding the possibility of packet losses due to collisions.

Figure 1 shows the density function of the corruption length in bytes of the Zigbee packets. The Zigbee packet length is 122 bytes. The density is always decreasing with the number of erroneous bytes. The observed density does not correspond to a geometrical distribution, which would be expected since the errors are expected to be independent. In fact, the errors in the sequence of the transmitted bits are independent but since spectrum spreading is used, when the received bit flow is de-spread, errors are cancelled (which is the reason why the spectrum spreading is efficient in low signal to noise plus interferences situations) resulting in a non exponential error distribution.

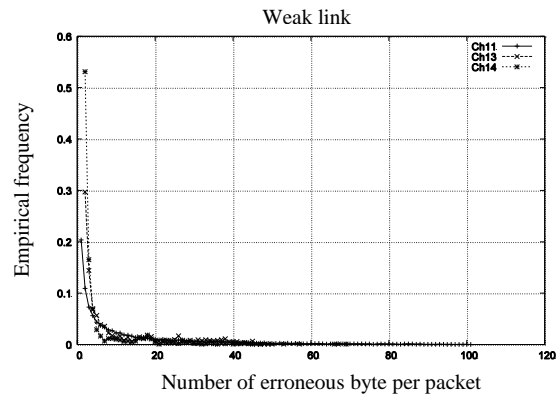

Figure 1: Empirical frequency of the error length for the weak link

- **ZigBee interference (Hidden terminal)**

In this scenario, one ZigBee transmitter sends traffic to its receiver while another ZigBee transmitter sends traffic to another one, both couples interfering together. One sender uses 122 byte packets. We varied the size of the packet sent by the other concurrent ZigBee sender (cf. Figure 2, Figure 3, Figure 4 and Figure 5). The Figure 2 represents the case where a couple communicates with packets of length 122 bytes while the other couple communicates with a packet length of 16 bytes. A peak is reached at 11 bytes with a density of 0.3. The Figure 3 represents the collision between 122 byte packets with 90 byte packets. A peak appears at 85 bytes with a density of 0.22. In the case of the Figure 4 all packets have an equal 122 byte length. Two peaks appear, at 1 byte and at 106 bytes. The Figure 5 represents collisions between 122 byte packets and packets of length 12 bytes.

A common feature between all these cases is the difference of 5 bytes between a packet and its collision effect on the 122 bytes packet. It is due to the fact that 5 is the length of the physical layer header. It is less clear in the case of collisions between 122 bytes packets and 12 bytes packets (cf. Figure 5). On Figure 4, the peak is at 106 bytes which is equal to 122-12(ZigBee header length)-4. The other peak at 1 byte is due to the big length of the packet and the Tx scheduling deviation which is related to the default swap between Tx and reception that exists in the CC2420 radio, the result is a collision with a small portion or a big portion of the packet. In the 122bytes long packets corruptions, the header size should be subtracted because the packets for which the header is corrupted are not received and taken in the analysis. In the case of the 122B vs.12B collision, due to the difficulty to have collision the dominating effect is a combination between noise and weak link effect.

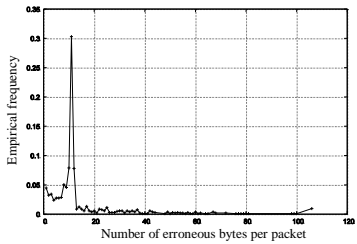

Figure 2: 122B vs. 16B packets

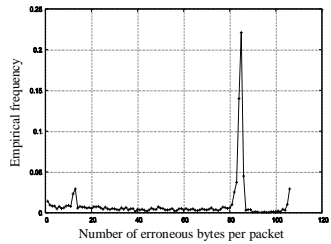

Figure 3: 122B vs. 90B packets

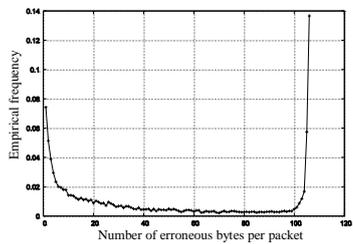

Figure 4: 122B vs. 122B packets

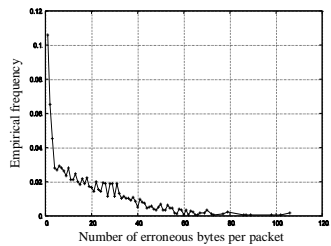

Figure 5: 122B vs. 12B packets

- **ZigBee with BlueTooth**

The next scenario (cf. Figure 6 to Figure 7) is intended to observe the Bluetooth fingerprint. During the test 68006 packets were transmitted of which 2531 packets were lost. The ZigBee rate and used channel are varied. For clarity and space purposes we only show the highest and lowest Zigbee Tx rates 12.5 packets per seconds (pps) and 166 pps. As it was expected, the coexistence with Bluetooth has a low effect on the Zigbee traffic. Less than 4% of the packets are lost due to collisions with Bluetooth. More precisely most of the lost packets were lost at the connection establishment between the Bluetooth devices (in this case the FH rate becomes 3200 hps).

Bluetooth has the same effect on the 3 channels and for the various rates used by Zigbee. The cause of the resemblance between the 3 channels is due to the frequency hopping performed by Bluetooth.

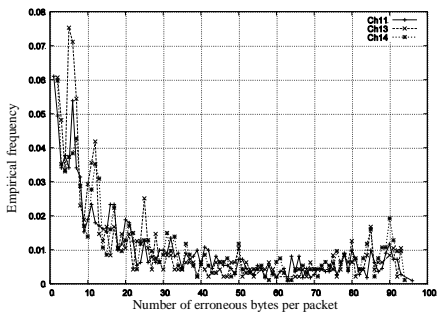

Figure 6: Zigbee colliding with BlueTooth, with a ZigBee Tx rate 12.5pps

Bluetooth is compared to the weak link case in Figure 8. In the Bluetooth graph most of the corruptions have densities less than 0.08, the same as the densities of corruption bigger than 4 bytes in the weak Link. The key differences are first the high density of the "one byte corruption" relatively to the other corruptions lengths in the weak SNR graph and second the fact that, at the tail of the Bluetooth density graph, there are higher densities that appear for the corruptions with big lengths, located between 80 bytes and 90 bytes, which are linear and flat in the case of the weak link.

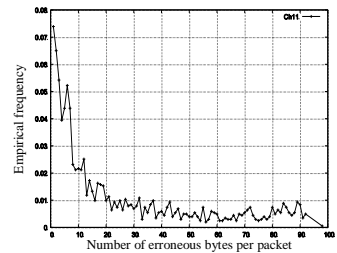

Figure 7: ZigBee colliding with BlueTooth, ( Zigbee Tx rate 166pps)

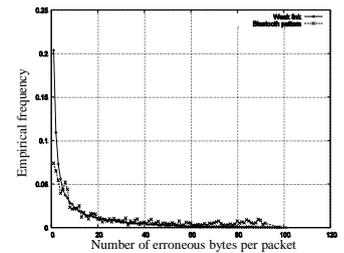

Figure 8: BlueTooth and weak link patterns(for comparison)

- **ZigBee with WiFi**

We should emphasize that the transmission rates used in the experiments of WIFI coexistence are based on real scenarios. In the first scenario, the downlink interference in the areas covered by the WIFI AP is much higher than uplink interference of the mobile WIFI nodes; the uplink is used for small request messages, in contrast the downlink generates a high aggregated traffic of different types and for different mobile nodes. In the second scenario, what will affect the sensor is the sum of all the traffic generated by that AP which will be very high if there were a high number of mobile nodes or big files transfer.

First we observe the effect of the control traffic only on different channels. The results are shown on the Figure 9, Figure 10 and the Figure 11. The overlap of a WiFi channel with Zigbee channels (cf. Figure 15) has various effects on ZigBee depending on the deviation from the center frequency of the WiFi channel. We varied the used channel but we only show the results of the WiFi channel 1 overlapped with the Zigbee channels 11, 13 and 14. The results can be extended to the other cases like the one with the WiFi channel number 1 together with the Zigbee channel 11 (similar to the effect of WiFi channel 7 on the Zigbee channel 17). On the Figure 9 and Figure 11, there are peaks at 26 and 27 bytes respectively. These peaks are caused by the collisions between ZigBee and the WiFi control packets (beacons, requests, etc.).

On Figure 10 the effect of WiFi is more resembling to the weak link case. This is due to the energy distribution of the WiFi traffic which is not uniform over its spectrum but more important on the left of the WiFi spectrum (cf. [10]). The used Zigbee Tx rate is 33 pps. During all the tests we did not modify the default control message Tx rate and length. For the channel 11, 51182 Zigbee packets are sent, 7% are lost. For the channel 13, 26444 packets are sent, 18% are lost. For the channel 14, 26167 packets are sent, 6% are lost.

Then, the WiFi data traffic is added (cf. Figure 12, Figure 13 and Figure 14). The data Tx rate is 916 pps with a 1500 byte packet length, the Zigbee rate is 166pps with a 122 byte packet length. On the channel 11, 51200 packets were transmitted and 6% were lost. On the channel 13, 51049 packets were sent and 18% were lost. On the channel 14, 51182 packets were sent and 0.7% were lost. The effect of the WiFi packets on the channel 11 is less than on the channel 13 due to its higher deviation from the WiFi center frequency. On the channel 13, there are two main peaks at 27 bytes and at 4 bytes, on the channel 11 there are similar peaks but the first one at a smaller value.

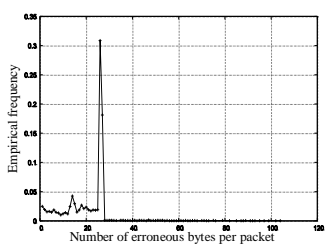
Figure 9 WiFi control traffic pattern, channel 11

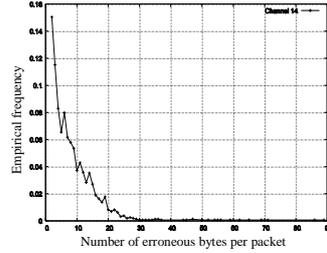
Figure 10 WiFi control traffic pattern, channel 14

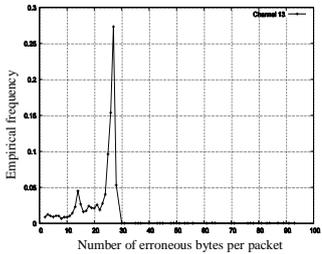
Figure 11 WiFi control traffic pattern, channel 13

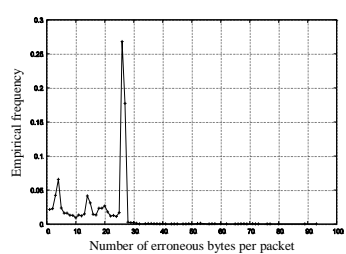
Figure 12: WiFi control and data traffic pattern, channel 11

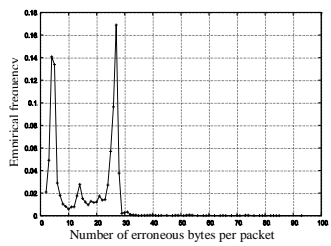
Figure 13: WiFi control and data traffic pattern, channel 13

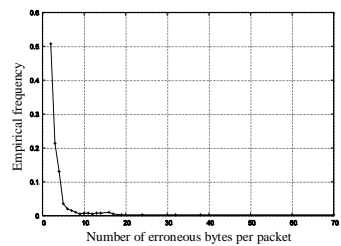
Figure 14: WiFi control and data traffic pattern, channel 14

The case of the channel 14 has the form of weak link case for the same reason as in Figure 10. The cause of the low packet loss percentage is that as the control packets use different modulation from the data packets it creates higher interferences on most of the wifi channel band, specifically the higher edge of the band where channel 14 of the Zigbee is located. At the same time the control packets use basic rate which implies that the channel occupation of these packets takes longer time than the data packets, this gives higher probability for collisions with Zigbee packets. During the experiment when we added data traffic to the control packets, this caused the transmitter to send less control packets on the expense of the data traffic. With less control packets sent less collisions happens. Therefore the dominant effect becomes more or less the weak link effect.

Finally, the WiFi has a fingerprint with a shape of a horse saddle. The first peak is related to corruptions caused by the data packets. The data packets cause short corruption bursts because of the high WiFi data bitrate and the fact that the ZigBee channels intersect only a small part of the WiFi one. The second peak is related to the control packets which use the basic rate which is different (2Mbps, cf. [10]) from the bit rate used by the data packets.

There is a small probability to have the case of overlapping APs that uses the same channels. Therefore its effect will be minimal on the fingerprint. Usually if there was adjacent APs they will use different channels to avoid inter cell interference. This act will cause the Zigbee to just interfere with one AP. As a result the fingerprint will not be affected.

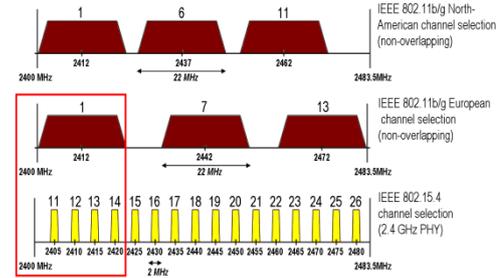
Figure 15: the red rectangular zone represents the experiment's channels used

Depending on the data traffic rate, the weight of each peak is different (cf. Figure 16), but the pattern does not change. The effect of the control packets (peak at 26-27bytes) is observed on the channels 11 and 13 but not clear on the ch 14.

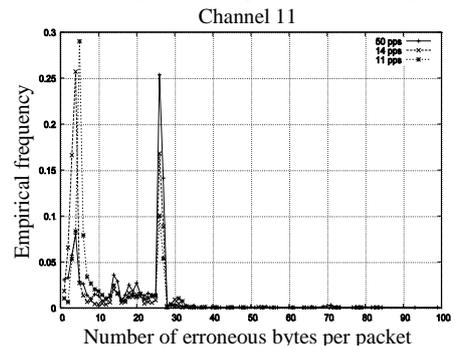
Figure 16: Wifi control and data traffic pattern, effect of the data rate

Experimentations with different WiFi packet lengths show that below a certain threshold packet length, equal to L=1100 bytes, the data peak disappears. In an experiment with a WiFi back to back packet transmission (1250 pps), the WiFi data traffic with 1100 byte length packets is still detectable (cf. Figure 17) but when a packet length of l=1000bytes, for example, is used, only the peak corresponding to the control traffic remains (cf. Figure 18). We do not represent the channel 14 in Figure 18 because we had only 0.2% packets lost over 39634 packets sent. For the channel 11, 1.5% packets are corrupted out of 38488 packets sent. For the channel 13, there are 19% corrupted packets out of 38672 packets.

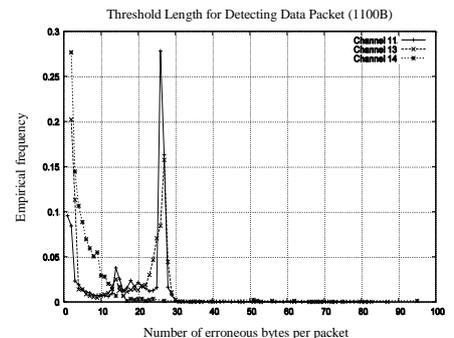
Figure 17: Wifi control and data traffic pattern, effect of the large packets

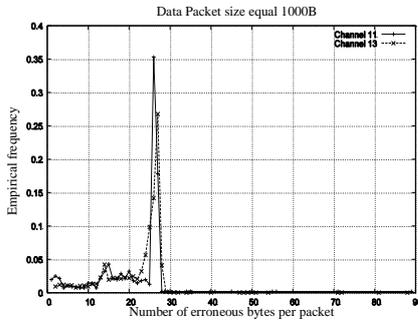

Figure 18: Wifi control and data traffic pattern, effect of small packets

- **ZigBee with a mix of concurrent networks**

In a last experiment, we mixed different types of concurrent technologies at the same time. We used a WiFi transmitter sending back-to-back 1400 byte long data packets to a receiver with, on the channel 11, two ZigBee transmitters sending traffic to a receiver, a ZigBee transmitter sending traffic to a receiver on the channel 13 and another ZigBee transmitter sending traffic to a receiver on the channel 14.

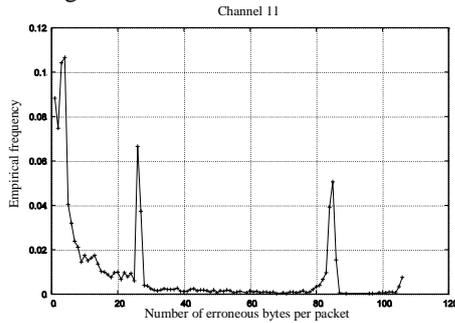

Figure 19: A mix of concurrent networks observed in channel 11

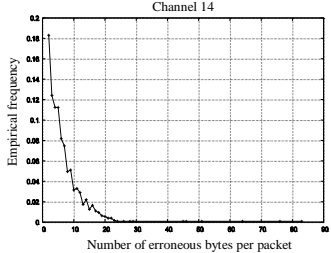 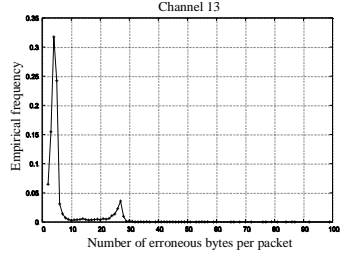

Figure 20: ZigBee with a mix of concurrent networks, channel 14

Figure 21: ZigBee with a mix of concurrent networks, channel 13

On the channel 11, a ZigBee node sends 122 byte long packets and the other 90 byte packets. The results are shown Figure 19, Figure 20, and Figure 21. The distinction between these technologies is clear. On Figure 19 representing the channel 11 the three peaks appear. The first peak, between 2 and 6 bytes, is due to the WiFi data packet. The second peak is due to the WiFi control packets and the last one at 85 bytes corresponds to the Zigbee packets. The zigbee channels being quasi-orthogonal, no zigbee effect is observed on the channels 13 and 14 (Figure 21, Figure 20 respectively). The wifi effect appears clearly on the channels 11 and 13.

Overlapping areas between APs can exist which gives the impression that the error patterns can be affected. In reality these areas are undesirable and are aimed to be small. In an industrial environment they prevent such interference by assigning different channels to each adjacent AP. The impact on the sensors is that it will only be interfered by one AP most of the time. Therefore the fingerprint of the WIFI should stay applicable most of the time in the area covered by the AP even if there were multiple WIFI mobile nodes. Furthermore if there were different types of traffic, the sensor will be affected mostly by the longest burst of data and by the length of the packets as it was shown in the experiments.

## IV. DISCUSSION

Although in reality it makes sense that one can imagine a high number of error patterns due to the chaotic nature of the interference. We aim to detect specific technologies of interest. The fact is that each technology uses specific components: type of modulation, Tx power and scheduling mechanism. For each component specific BER is demonstrated in the literature. This suggests that the combination of such component in a specific technology will create a unique combination of BER. This is shown in the experiments and presented as error patterns in a received packet. Moreover the experiments showed that due to the use of a number of packet not just one sample, the patterns converge to the dominating interferer fingerprint.

The remarkable point is that the case of coexistence will always be detected even if the concurrent technology has a small overlapped portion of the ZigBee spectrum, or small duration of occupying the spectrum or low power. It is deduced from the experiments (cf. figure 5, 6, 7, 10, 14 and 20). In all of these figures a dominant peak at the one byte along with a wide base that represents the high error density at high error lengths that reaches 20 bytes is shown. This is what represents a distinction from the weak link effect. Although in this case we cannot identify a specific technology but we can be sure that the node is in the case of coexistence and the node should react accordingly.

## V. FIM AND ITS APPLICATION TO LINK ADAPTATION

From the above results, we designed a Fingerprint Identification Mechanism, FIM, allowing to recognize on the fly the technology used in the concurrent network. Our observations showed that the ability to determine the cause of packet corruptions from the number of erroneous bytes is highly accurate and simple. There is no need to use sophisticated costly methods for the detection. Each corrupted packet is stored in a queue (in push out mode because of the limited memory) and a negative acknowledgment is sent. Each time a correct packet (correct CRC) is received a search for the packets with the same ID and the sequence number is done in the queue. Then the number of erroneous bytes in the corrupted packets is calculated and an empirical frequency of the number of erroneous bytes is updated. Then a comparison is triggered to detect peaks at certain error length. When a fingerprint is detected, an action is triggered. Otherwise if there was an ambiguity in the pattern, FIM would not react.

To demonstrate the efficiency of our proposal, we implemented FIM with a simple CSMA/CA (we reactivated the disabled feature; the CCA, backoff, etc.), in a coexistence environment with WiFi. Two scenarios have been studied:

without and then with FIM to detect WiFi. In this last scenario, a WiFi detection triggers a channel swap from the channel 11 to the channel 15 in order to avoid the WiFi. This approach is only to serve as a prototype implementation of FIM and is by no means an optimal algorithm.

The aim of this scenario is to demonstrate how FIM can be effective in making more intelligent decisions to optimize a link quality in a self organized manner. We made two experiments, one with a data transmission rates used by WiFi equal to 458 pps and for the other the rate is 916 pps, the packet length being always 1500 bytes. The Zigbee data rate is 33pps. During these experiments we used typical traffic rates for scenario like intrusion detection, cold chain monitoring or health monitoring, along to the scenario where a node is a critical node (bottleneck): cluster head or a relay node. The presence of such nodes is critical to create bridges between the clusters and the parts of a network. The links connecting such critical nodes usually generate a high traffic that aggregates different traffics passing through to the different parts of the network. If these nodes are lost the connectivity among the network nodes is lost. The rates in these circumstances can exceed 50 pps in case of an alert activation or for continuous surveillance.

After 105 collisions for 458 pps and 20 collisions for 916 pps the WiFi is detected. The Figure 23 shows the throughput change through time for the case 458 pps. At the link adaptation activation, that is to say at 173s, the change can be observed from 18 pps to 28 pps. On Figure 22, the average global improvement in throughput is shown. For a WiFi Tx rate of 458pps the gain is equal to 87%, and for 916 it is 100.9%. FIM is designed for fingerprint detection. It also indirectly gives an indicator of the effectiveness of the countermeasures used against the concurrent technologies. The more the countermeasure is effective the less erroneous bytes there should be and thus the less a concurrent technology is detected. Moreover, the efficiency of FIM increases with the collisions. If there are few collisions, it converges slowly, but in this case the problem of the collisions is less crucial: if the effect of the coexisting technologies is minimal or there is no effect there is no need to adapt.

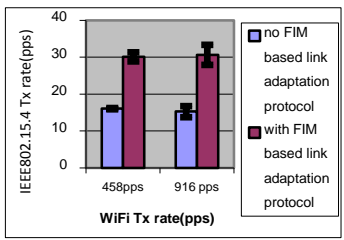

Figure 22: throughput with and without FIM for link adaptation

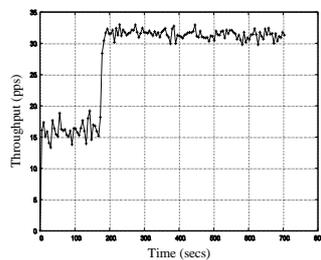

Figure 23: ZigBee throughput at the detection instant for a WiFi rate of 458pps and a Zigbee rate of 33.3pps

## VI. CONCLUSION

The diversity of technologies that coexist in the same ISM band as WSN, varies from technologies that do not perform listening before transmission like Bluetooth, to the ones that, due to their CCA specifications, do not detect other technologies like WIFI. Due to the lack of coordination between these technologies, the WSN needs to detect the presence of each one. Thus a specific countermeasure can be applied to avoid collisions with these technologies.

In this paper, we identified different fingerprints of various coexisting technologies based on the IEEE802.15.4 packet corruption patterns. We considered ZigBee, BlueTooth and WiFi concurrent traffics, each one producing a specific corruption pattern. Moreover the weak link corruption model is also identified. Then, we designed FIM, a reactive mechanism to detect on the fly these fingerprints. On our "Tmote Sky" testbed, in some cases FIM provides up to 100% accuracy in detecting the correct fingerprint. During some of the experiments, its implementation for link adaptation improved the throughput by 100.9%.

This preliminary work is to serve as a proof of concept. We performed first experiments which must be extended to other technologies and configurations, like with WiFi access points rather than in ad-hoc mode, using OFDM instead of DSSS, etc. The impact of the node mobility on the error shapes is also an interesting subject. In this paper we only give an implementation example of adaptation mechanism after WiFi detection to illustrate the efficiency and usefulness of our link adaptation mechanism. A lot of other adaptation mechanisms can be designed and optimized and for various other technologies.

ACKNOWLEDGMENT

The authors want to thank Prof. Monique BECKER and Dr. Vincent Gauthier for useful suggestions and comments.